\documentclass[11pt,twoside]{article}
\usepackage{asp2004}
\usepackage{epsfig,graphicx,natbib,url} 
\usepackage{thispreprint} 
\usepackage{lscape}
\bibpunct{(}{)}{;}{a}{}{,}    
\def\cite#1{\citealp{#1}}     

\newcount\longrefs
\def\aap{\ifnum\longrefs=1 {Astron.\ Astrophys.}\else 
                           {A\hbox{\rm \&}A}\fi}
\def\aapr{\ifnum\longrefs=1 {Astron.\ Astrophys.\ Rev.}\else 
                            {A\hbox{\rm \&}AR}\fi}
\def\aaps{\ifnum\longrefs=1 {Astron.\ Astrophys.\ Suppl.}\else 
                            {A\hbox{\rm \&}A Suppl.}\fi}
\def\aj{\ifnum\longrefs=1 {Astron.\ J.}\else 
                          {AJ}\fi} 
\def\ao{\ifnum\longrefs=1 {Applied Optics}\else 
                           {Appl.\ Opt.}\fi} 
\def\aspcs{\ifnum\longrefs=1 {Astron.\ Soc.\ Pacific Conf. Series}\else 
                           {ASP Conf.\ Ser.}\fi} 
\def\apj{\ifnum\longrefs=1 {Astrophys.\ J.}\else 
                           {ApJ}\fi} 
\def\apjl{\ifnum\longrefs=1 {Astrophys.\ J. Lett.}\else 
                            {ApJ}\fi} 
\def\aplett{\ifnum\longrefs=1 {Astrophys.\ J. Lett.}\else 
                            {ApJ}\fi} 
\def\apjs{\ifnum\longrefs=1 {Astrophys.\ J. Suppl.}\else 
                            {ApJS}\fi}
\def\apss{\ifnum\longrefs=1 {Astrophys.\ and Space Science}\else 
                            {Astrophys.\ Space Sci.}\fi}
\def\araa{\ifnum\longrefs=1 {Ann.\ Rev.\ Astron.\ Astrophys.}\else 
                            {ARA\hbox{\rm \&}A}\fi}
\def\azh{\ifnum\longrefs=1 {Astronomicheskii Zhurnal}\else 
                            {Astron.\ Zhur.}\fi}
\def\baas{\ifnum\longrefs=1 {Bull.\ Am.\ Astron.\ Soc.}\else 
                            {BAAS}\fi}
\def\bain{\ifnum\longrefs=1 {Bull.\ Astronom.\ Institutes Netherlands}\else
                            {Bull.\ Astr.\ Inst.\ Neth.}\fi}
\def\gca{\ifnum\longrefs=1 {Geochim.\ Cosmochim.\ Acta}\else 
                           {Geochim.\ Cosmochim.\ Acta}\fi}
\def\grl{\ifnum\longrefs=1 {Geophys.\ Res.\ Lett.}\else 
                           {Geoph.\ Res.\ Lett.}\fi}
\def\iaucirc{\ifnum\longrefs=1 {IAU Circulars}\else 
                          {IAU Circ.}\fi}
\def\ip{\ifnum\longrefs=1 {in press}\else 
                          {in press}\fi}
\def\jgr{\ifnum\longrefs=1 {J.\ Geophys.\ Res.}\else 
                           {J.\ Geophys.\ Res.}\fi}  
\def\jrasc{\ifnum\longrefs=1 {J.\ Royal Astron.\ Soc.\ Canada}\else 
                           {JRAS Can.}\fi}  
\def\mnras{\ifnum\longrefs=1 {Mon.\ Not.\ Roy.\ Astron.\ Soc.}\else 
                             {MNRAS}\fi} 
\def\nat{\ifnum\longrefs=1 {Nature}\else 
                           {Nat}\fi}
\def\pasj{\ifnum\longrefs=1 {Pub.\ Astron.\ Soc.\ Japan}\else 
                            {PASJ}\fi} 
\def\pasp{\ifnum\longrefs=1 {Pub.\ Astron.\ Soc.\ Pacific}\else 
                            {PASP}\fi} 
\def\physscr{\ifnum\longrefs=1 {Physica Scripta}\else 
                            {Phys.\ Scrip.}\fi} 
\def\planss{\ifnum\longrefs=1 {Planetary \& Space Science}\else 
                            {Plan. \& Space Sci.}\fi} 
\def\procspie{\ifnum\longrefs=1 {Proc.\ SPIE}\else 
                            {Proc.\ SPIE}\fi} 
\def\qjras{\ifnum\longrefs=1 {Quarterly J.\ Royal Astron.\ Soc.}\else 
                            {QJRAS}\fi} 
\def\sa{\ifnum\longrefs=1 {Soviet Astron..}\else 
                               {Sov.\ Astron.}\fi}
\def\skytel{\ifnum\longrefs=1 {Sky \& Telescope}\else 
                            {Sky \& Tel.}\fi} 
\def\solphys{\ifnum\longrefs=1 {Solar Phys.}\else 
                               {Sol.\ Phys.}\fi}
\def\ssr{\ifnum\longrefs=1 {Space Science Rev.}\else 
                               {Space\ Sci.\ Rev.}\fi}

\def\cf{cf.}                       

\def\specchar#1{\uppercase{#1}}    

\def\CaII{\mbox{Ca\,\specchar{ii}}}

\def\HI{\mbox{H\,\specchar{i}}} 
 
\def\HeI{\mbox{He\,\specchar{i}}}

\def\Halpha{\mbox{H\hspace{0.1ex}$\alpha$}} 

\def\Hbeta{\mbox{H\hspace{0.2ex}$\beta$}}

\def\Lyalpha{\mbox{Ly$\hspace{0.2ex}\alpha$}}

\def\CaIIK{\mbox{Ca\,\specchar{ii}\,\,K}}       
\def\CaIIH{\mbox{Ca\,\specchar{ii}\,\,H}}

\def\HK{\mbox{H\,\&\,K}}
\def\Hthree{\mbox{H$_3$}}

\def\HtwoV{\mbox{H$_{2V}$}}

 \def\rmA{{\rm A}}             

\def\rms{{\rm s}}

\def\is{\!=\!}                             
\def\ep{\:{\rm e}^}                        

\def\={\hbox{$\!=\!$}}                     

\begin{document}

\def\rrtitle{On the Nature of the Solar Chromosphere} 
\markboth{Robert J. Rutten}{\rrtitle}   
\title{\rrtitle}
\author{Robert J. Rutten}
\affil{Sterrekundig Instituut, Utrecht University, The Netherlands \\
       Institute of Theoretical Astrophysics, University of Oslo, Norway}

\begin{abstract} 
  DOT high-resolution imagery suggests that only internetwork-spanning
  \Halpha\ ``mottles'' constitute the quiet-sun chromosphere, whereas
  more upright network ``straws'' in ``hedge rows'' reflect
  transition-region conditions.
\end{abstract}

\section{Introduction}
\begin{figure}
  \centering
  \includegraphics[width=\textwidth]{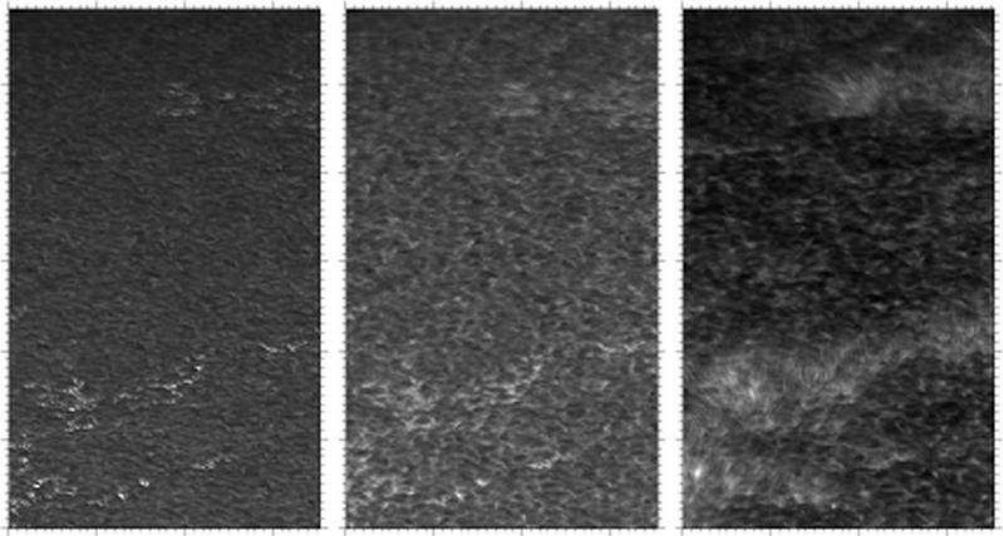}
  \caption[]{\label{fig:limb-images}
  Three partial near-limb images taken on June 18, 2003.  Ticks at
  arcsec intervals.  The limb is just above the top.  The first two
  are simultaneous, the third was taken 25\,s earlier.  {\em Left\/}:
  G band, showing the onset of reversed granulation and short bright
  stalks where our view penetrates through relatively empty network
  fluxtubes into hot granules.  {\em Center\/}: \CaIIH\ wing, a
  similar scene sampled slightly higher up.  {\em Right\/}: \CaIIH\
  line center.  The dark background consists of reversed granulation
  sampled by the inner-wing contributions within the 1.4\,\AA\ FWHM
  passband wherever the \CaIIH\ core contributes insignificant
  magnetic-feature emission.  Some internetwork features may mark
  acoustic shocks (but \HtwoV\ ``grain'' amplification from \Hthree\
  blueshift may lack along these slanted lines of sight).  The active
  network is characterized by crowded forests of long thin bright
  features contributed by the line core.  They are obviously optically
  thin.  They are mostly upright since the internetwork foreground
  remains dark out to their bottoms.  They start close to the
  photospheric network bright points.  They are very dynamic
  (\url{http://dot.astro.uu.nl/movies/2003-06-18-mu034-ca-core.mpg}).
  I call them ``straws'' but my Utrecht colleagues prefer ``grass''.
}\end{figure}

\begin{figure}
  \centering
  \includegraphics[width=\textwidth]{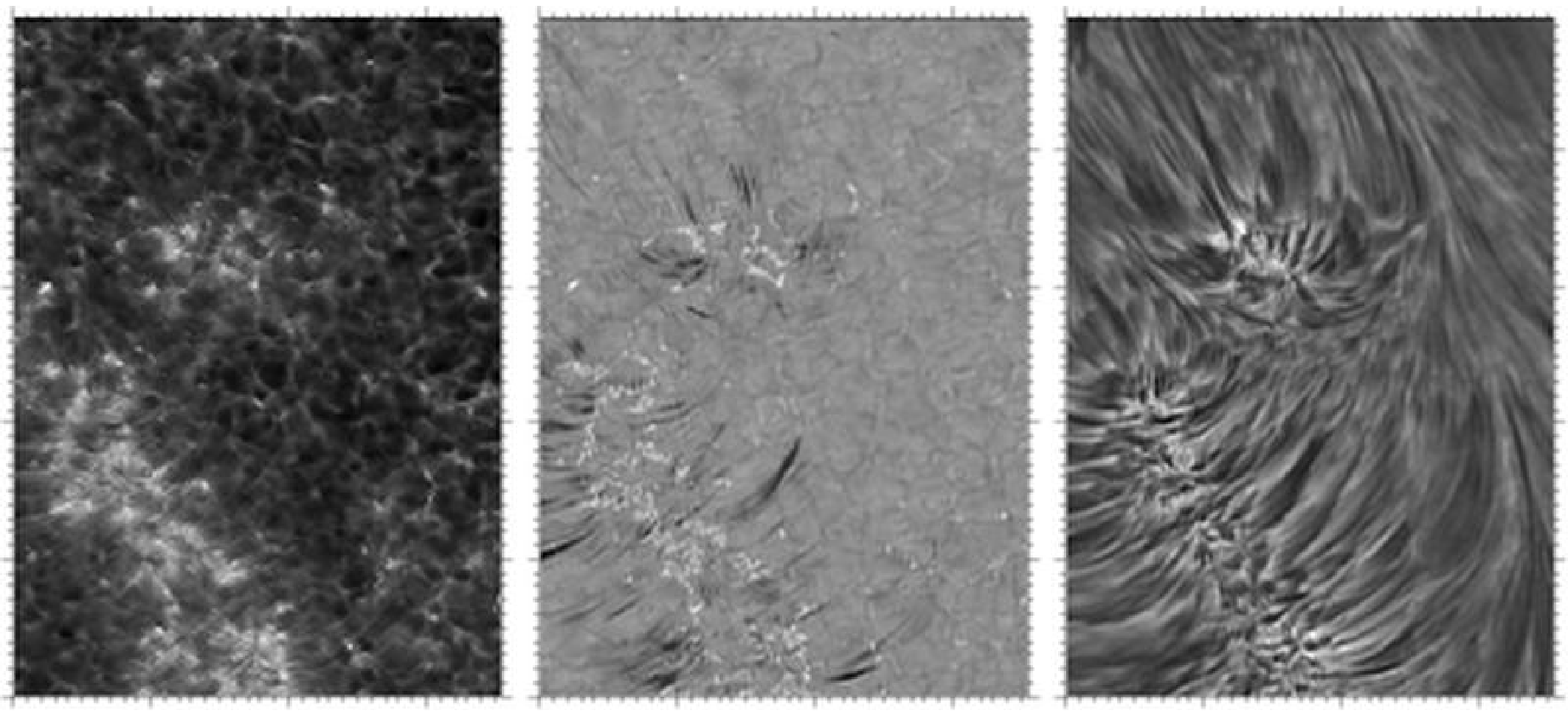}
  \caption[]{\label{fig:disk-images}
  Three partial disk-center images taken simultaneously on October 6,
  2004.  Ticks at arcsec intervals.  A filament lay along the
  righthand side of this cutout
  (\url{http://dot.astro.uu.nl/movies/2004-10-06-filament-4p.mpg}).
  {\em Left\/}: \CaIIH.  The internetwork shows reversed granulation
  contributed by the inner wings, superimposed acoustic \HtwoV\
  grains, and a few isolated magnetic elements (``persistent
  flashers''). The network bright points are less sharp then and
  differ in morphology from G-band bright points.  Bright stalks
  emanate a few arcsec from them at this resolution (zoom in with your
  pdf viewer).  Diffuse \CaIIH\ core brightness spreads as far or
  further.  {\em Center\/}: blue \Halpha\ wing at 800\,m\AA\ from line
  center.  The internetwork consists of normal granulation at very low
  contrast.  Photospheric magnetic elements appear very bright for
  various reasons explained by
    \citet{Leenaarts++2006a}. 
%
%
%
  Their morphology is identical to G-band bright points although they
  are less sharp at given telescope resolution.  The very dark streaks
  that emanate from the network, often starting at some distance,
  frequently appear so dark through blueshift.  Their abrupt outer
  ends vary rapidly in location.  {\em Right\/:} \Halpha\ line center.
  A mass of extended mottles.  They curve away from the neutral line
  under the filament.  They appear intransparent.  Their frequently
  bright beginnings are near to, but not co-spatial with, yet brighter
  network grains that have a closer resemblance to the \CaIIH\ ones
  than to the \Halpha-wing ones.
}\end{figure}

In my talk I skipped a planned discourse on definitions of the
chromosphere because Philip Judge had just admirably done so (these
proceedings).  I then used movies from the Dutch Open
Telescope\footnote{All DOT
  data are public and available under {\em DOT database} at
  \url{http://dot.astro.uu.nl}.  The movies that I showed and sample
  here in Figs.~\ref{fig:limb-images} and \ref{fig:disk-images} reside
  there also, under {\em DOT movies}.}
  (\cite{2004A&A...413.1183R}) 
to argue that much so-called chromosphere is actually transition
region, that much other so-called chromosphere is actually upper
photosphere, and that only those \Halpha\ mottles and fibrils that
span across and between internetwork cells and active regions
constitute the actual chromosphere.  I compress the argument here into
a two-figure summary of the evidence and a list of brief conjectures
with explanatory cartoons.  The upshot is that radiation modeling of
the field-guided filamentary features (mottles, fibrils, spicules,
``straws'') that infest -- if not constitute -- the quiet-sun domain
between photosphere and corona should mix steep gradients in varied
configurations with thick-to-thin radiation simulation containing
everything from LTE to coronal conditions.  One conjecture is that the
latter dominate such things as \CaII\ \HK\ superbasal
magnetic-activity emissivity, making that a transition-region
diagnostic.

\section{Evidence}
Fig.~\ref{fig:limb-images} shows \CaIIH\ ``straws'' at right: long
thin emission features in \CaIIH, in rows jutting out from active
network, seen near the limb against a dark background of reversed
granulation.  The latter is photospheric and is sampled only slightly
deeper in the other two images.

Fig.~\ref{fig:disk-images} shows active network on the disk.  The
strikingly bright \Halpha-wing\ network bright points form in the deep
photosphere; the different-morphology \CaIIH\ network bright points
form higher up.  The \Halpha\ core shows its proprietary filamentary
structures habitually called ``mottles'' in quiet sun, ``fibrils''
when jutting out from active regions, ``spicules'' off the limb.
Those in Fig.~\ref{fig:disk-images} are mottles but with large-scale
organization from avoiding the filament (the campaign target) just
outside these cutouts.  They tend to have bright near-network
beginnings, especially in active plage. 

The \Halpha\ disk mottles are not seen in \CaIIH.  Reversely, \CaIIH\
limb straws do not stand out between all other mottles seen in
\Halpha\ line center near the limb.  They gain hedge-row prominence
there only in the outer \Halpha\ wings, as slender isolated dark
features against the bright \Halpha-wing background which appears when the
internetwork is no longer occulted by cell-spanning mottles.

The VAULT-2 \Lyalpha\ images at
\url{http://wwwsolar.nrl.navy.mil/rockets/vault/archives.html}
  (\cf\ \cite{2001SoPh..200...63K}) 
show dense, 5000\,km-high hedge rows of abruptly ending bright upright
straws above network near the limb, twice as high and much thicker
than in \CaIIH.  In addition, they show weaker and flatter rosettes
fanning out from network that also end abruptly, an opaque dark floor
of long mottles covering internetwork areas.  On-disk plage appears as
a dense forest of short bright stalks and bright grains.

Comparison of the \CaIIH, \Halpha, and \Lyalpha\ scenes raises
questions:
\leftmargini=3ex
\begin{enumerate} \itemsep=0ex \vspace*{-2ex}

  \item Similarities: how can these various structures appear
        simultaneously in \Lyalpha, \Halpha, and \CaIIH?  A similar
        question as the old one why off-limb spicules appear in all of
        \HeI\ D$_3$, \Halpha, and \CaIIK, but adding \Lyalpha.

  \item Differences: the hedge rows of upright straws are bright and
        optically thick in \Lyalpha, bright and thin in \CaIIH, much
        less distinct and dark in \Halpha\ line center, more prominent
        but less upright and very dark in the \Halpha\ wings.  The
        cell-spanning \Halpha\ mottles are not seen in \CaIIH\ and
        only as a dark floor in \Lyalpha.  Why?

\end{enumerate}

\section{Considerations}
\begin{figure}
  \centering
  \includegraphics[width=4cm]{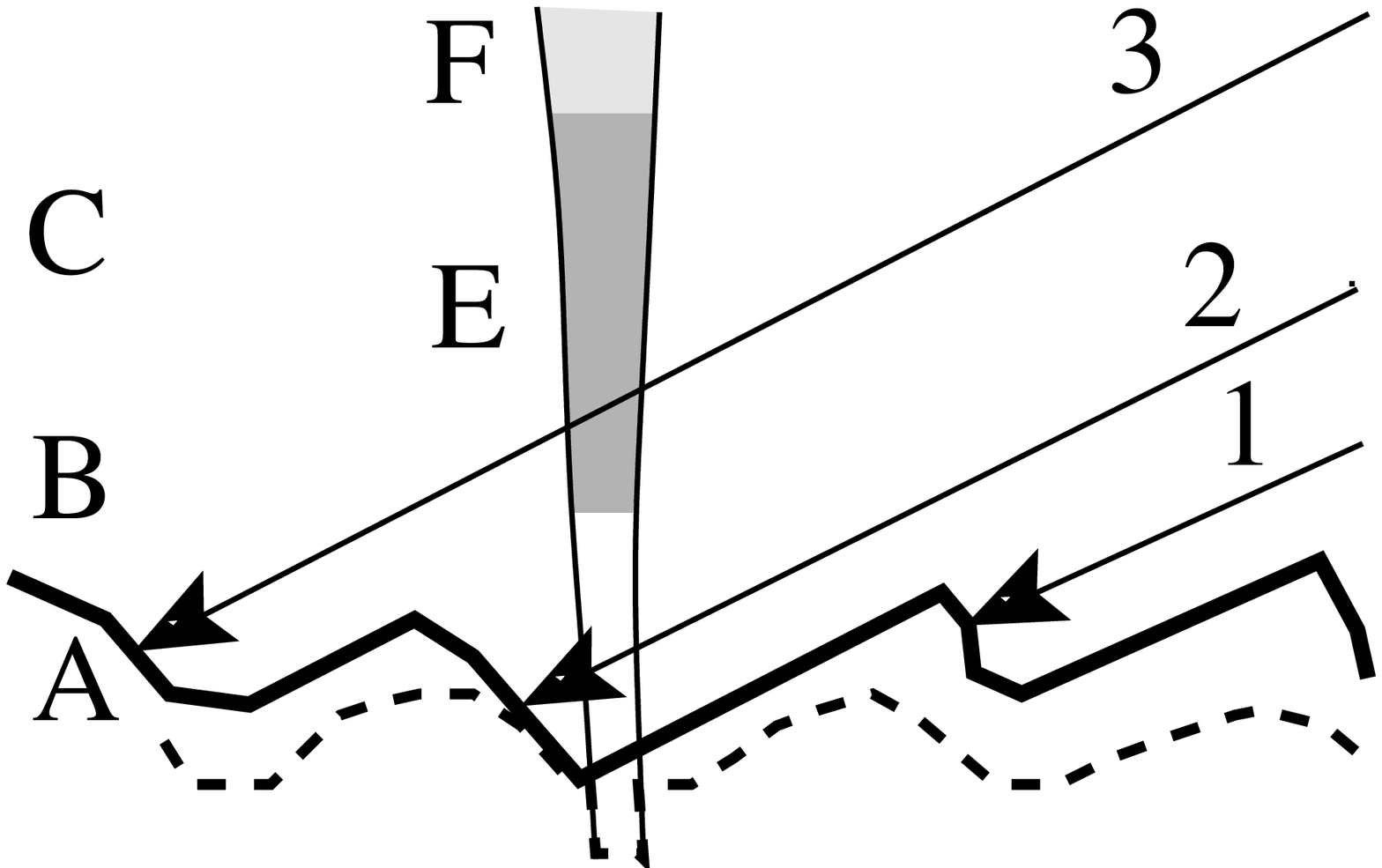}
  \hspace{1cm}
  \includegraphics[width=7cm]{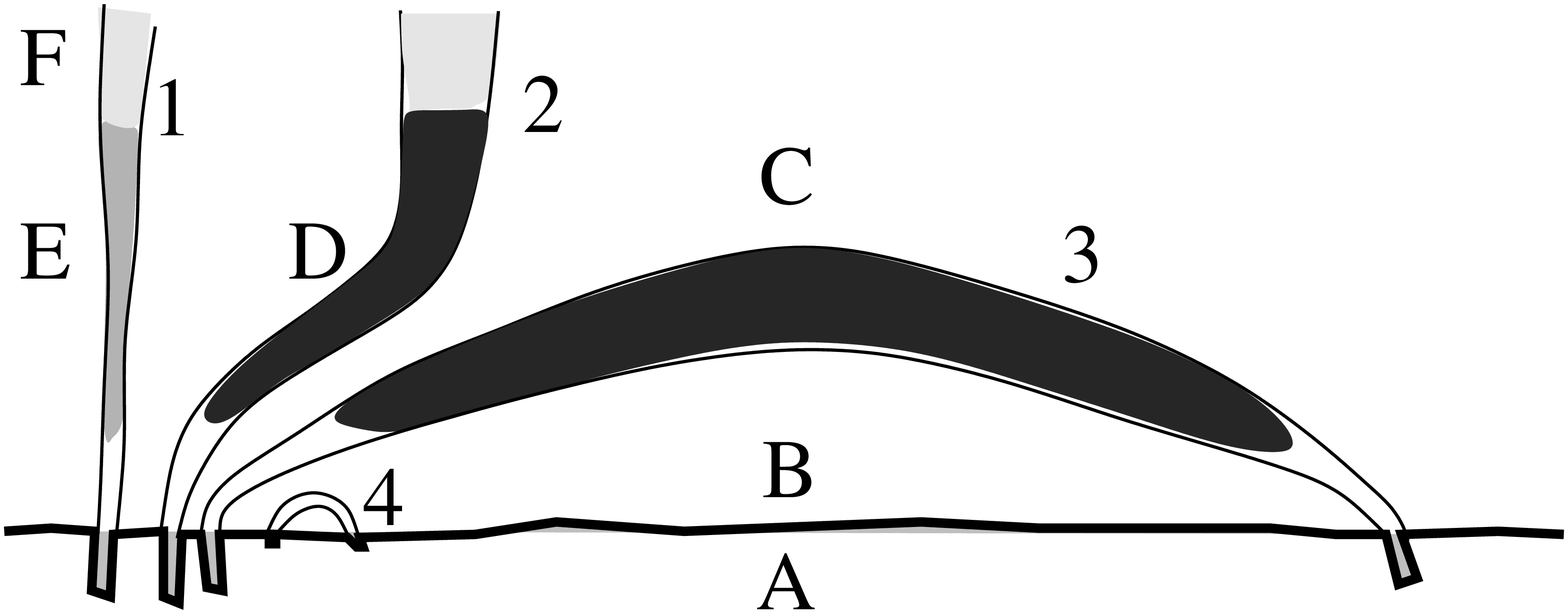}
  \caption[]{\label{fig:cartoons}
  {\em Left\/}: near-limb \CaIIH\ observation.  Line of sight 1 hits
  $\tau \is 1$ (thick curve along the bottom, the dashed curve is
  $\tau \is 1$ for radial viewing) in a granule and produces low
  brightness $I \approx B\,(\tau \is 1)$ in the inner wings of \CaIIH\
  which sample reversed granulation in the middle photosphere (regime
  A).  It passes first through internetwork regime C which is fully
  transparent in \CaIIH.  The contribution of line-center emission by
  acoustic shocks in upper-photosphere regime B is also small.  Line 2
  hits another granule after passing through a fluxtube that is
  relatively empty at low height.  The smaller opacity buildup results
  in deeper sampling than along lines parallel to line 2 in the plane
  perpendicular to the sketch but missing the fluxtube similarly to
  line 1. The granular interior therefore appears as a short bright
  stalk in the \CaIIH\ wings.  Line of sight 3 also hits a granule
  after passing through the fluxtube, but higher up through regime E
  which contributes the straw brightness.\\
  {\em Right\/}: different filamentary structures.  A similar cartoon
  is shown in Fig.~2 of
  \citet{1993ApJ...406..319F}; 
  compare also Judge's sketch in these proceedings.  Rough
  temperatures: D$\sim$10$^4$\,K, E$\sim$10$^5$\,K, F$\sim$10$^6$\,K. 
  Outside regimes: A = photosphere with
  normal/reversed granulation and tenuous magnetic elements, B = upper
  internetwork photosphere pervaded by acoustic shocks but
  nevertheless cool, C = transparent in \HK, \Halpha, and \Lyalpha.
  {\em Type 1\/}: bright upright \HK\ and \Lyalpha\ network straws
  opening into coronal plasma.  On the disk they produce grainy \HK\
  line-center, \Halpha\ line-center, and \Lyalpha\ full-profile
  emission in network and plage (Fig.~\ref{fig:disk-images}).
  {\em Type 2\/}: dark \Halpha\ mottles bending upward into hot plasma
  from unipolar crowding.  They represent the on-disk ``spicules''
  modeled by
  \citet{2004Natur.430..536D} 
  as due to $p$-mode mass loading. They end abruptly where the shocked
  cool gas meets transition-region temperatures.  Above plage they
  cause rapid occultive flickering of TRACE 171\,\AA\ brightness
  (``moss'') through bound-free hydrogen and helium scattering out of
  the TRACE passband
  (\cite{1999SoPh..190..409B}, 
   \cite{1999ASPC..184..181R}). 
  {\em Type 3\/}: dark \Halpha\ mottles spanning across cell interiors
  in bipolar network without hot-plasma connectivity.  They are
  reasonably well represented by the classical modeling of
  \citet{1967AuJPh..20...81G}, 
  Beckers (1968, 1972),
    \nocite{1968SoPh....3..367B} 
    \nocite{1972ARA&A..10...73B} 
  \citet{1994A&A...282..939H} 
  as opaque clouds of order 10$^4$\,K.
  {\em Type 4\/}: short weak-field near-network loops postulated by
  \citet{2003ApJ...597L.165S}. 
  They lack the Wilson depressions that turns types 1--3 into bright
  points in photospheric diagnostics and the mass loading that makes
  types 2 and 3 dark in chromospheric \Halpha.
}\end{figure}

The left part of Fig.~\ref{fig:cartoons} is a schematic consideration
of the scenes in Fig.~\ref{fig:limb-images}.  Mid-photosphere regime A
(tenuous fluxtubes in reversed granulation) is currently well
understood: time-dependent magnetoconvection plus LTE radiation
simulations explain G-band bright points, faculae, and reversed
granulation very well
  (\cite{2004A&A...427..335S}, 
   \cite{2004ApJ...607L..59K}, 
   \cite{2004ApJ...610L.137C}, 
   \cite{2005A&A...431..687L}). 
Internetwork upper-photosphere regime B requires time-dependent
NLTE modeling as in the acoustic shock simulation of
  \citet{1997ApJ...481..500C}, 
but does not produce noticeable \CaIIH\ emission outside acoustic
grains because the shocks do not heat on average.  Internetwork regime
C is transparent in \CaIIH.  Fluxtube regime E makes the straws in
Fig.~\ref{fig:limb-images} bright.  Fluxtube regime F starts where
they end, corresponding to spicule heights on the limb.

Since the straws appear without contributions from regimes B and C 
their intensity is, in homogeneous-cloud approximation:
\begin{equation}
  I \approx B(T_\rmA) \, \ep{-\tau_\rms}
        + S_\rms \, [1 - \ep{-\tau_\rms}] 
    \approx  
        B(T_\rmA) 
        + [S_\rms -  B(T_\rmA)]\, \tau_\rms,
  \label{eq:I}
\end{equation}
where $T_\rmA$ is the $\tau=1$ background temperature and $\tau_\rms$
the optical thickness of the straw along the line of sight.  The
second version requires $\tau_\rms \ll 1$. The straw source function is:
\begin{equation}
  S_\rms 
       = [(1-\varepsilon)\,\overline{J} 
           + \varepsilon\,B(T_\rms) 
           + \eta\,B^\ast]\,/\,(1+\varepsilon+\eta) 
           \approx (b_u/b_l)\,B(T_\rms),
  \label{eq:S}
\end{equation}
adding the contributions from resonant scattering, thermal
lower-to-upper level excitation, and multi-level roundabout photon
production.  In the optically thin limit without irradiation (as in
coronal conditions) use of the emissivity and geometrical straw
thickness $D$ is more direct: $ I \approx S_\rms \, \tau_\rms = j_\rms
\, D = (h\nu/4\pi) \, n_u \,A_{ul} \, D $.

\section{Conjectures}
\leftmargini=3ex
\begin{enumerate} \itemsep=0ex

  \item Structures: there are four distinct types of filamentary
        structures in the quiet-sun regime between photosphere and
        corona, sketched at right in Fig.~\ref{fig:cartoons}.  Type 1
        and 2 connect upward into hot conditions where they end
        abruptly.  Type 1 is rather straight and upright, remains
        slender to large height, and is hot and dense from
        unidentified mass and energy filling.  Type 2, often arranged
        in sheets, is less hot from propagative $p$-mode mass loading
        from below as suggested by De Pontieu et al.  Type 3 is
        supposedly loaded similarly or by siphoning, but does not
        connect into hot conditions and remains cool.  Type 4 is
        conjectured by
          \citet{2003ApJ...597L.165S} 
        but is hard to observe if neither evacuation nor filling gives
        them brightness signature.

  \item Regimes: the regime C surroundings of types 1--3 are
        transparent in \CaIIH, \Halpha, and \Lyalpha, most likely from
        being coronally hot and tenuous.  Upper-photosphere regime B
        is cool outside shocks, does not produce much \CaIIH\
        line-center emissivity, is transparent throughout \Halpha\ due
        to low \HI\ $n \is 2$ population, and is largely shielded by
        type-3 loops in \Lyalpha\ and \Halpha\ center.

  \item \Lyalpha: type 1 produces the thick hedge rows of network
        straws in the VAULT-2 near-limb images.  Their appearance
        suggests thermal \Lyalpha\ emission from optically thick
        straws with $I \approx S_\rms \approx
        \varepsilon_\rms\,B(T_\rms)$ if effectively thin and growing
        to $\sqrt{\varepsilon_\rms}\,B(T_\rms)$ for the thickest ones.
        Since $\varepsilon_\rms$ is likely small, $T_\rms$ must be high.
        The bright grains abounding in disk plage are due to type-1
        along-the-straw viewing.  The abruptly ending rosette fans
        correspond to type 2.  Type 3 constitutes the dark opaque
        internetwork background, with $I \approx \sqrt{\varepsilon_3}
        \, B(T_3)$.  Regime C must be transparent through being
        coronal, not only hot but also tenuous since at coronal
        temperatures \HI\ maintains as much as $10^{-6}$ fractional
        population (J.~Raymond).

  \item Other UV lines: type 1 produces network brightness in UV lines
        across many stages of ionization.  The systematic downdrafts
        observed in these suggest energy transfer from above.
        
  \item \CaIIH: type 1 also produces the \CaII\ near-limb straws, in
        similar fashion to the \Lyalpha\ ones but at smaller opacity,
        half as long, and with photospheric internetwork background.
        The $\eta B^\star$ contribution in Eq.\,(\ref{eq:S}) is likely
        important through dielectronic recombination since \CaII\
        ionizes to closed-shell Ar configuration.  It still has
        10$^{-3}$ fractional population at $T \is 10^5$\,K in coronal
        conditions
          (\cite{1998A&AS..133..403M}), 
        or nearly 1\% of the \HI\ $n \is 2$ population density.  In
        any case, \CaII\ level $n \is 2$ must have considerable
        population in the straws compared to regime C.  Somewhere
        along the fluxtube the magnetic-element brightness must flip
        from being due to evacuation to being due to excess filling
        (\cf\ \cite{2005A&A...437.1069S}). 
        On the disk, viewing along straws produces the bright \CaIIH\
        network grains and the short bright filaments jutting out from
        the \CaIIH\ network at high resolution
        (Fig.~\ref{fig:disk-images}).  The straws also cause the
        wide-spread diffuse \CaIIH\ brightness around network, the
        diffuseness arising from lack of resolution, small optical
        thickness through slanted straws, and resonance and electron
        scattering.

  \item \Halpha\ on the disk: type-3 mottles occult the underlying
        photosphere at line center.  They can be dark or bright
       (\cf\ \cite{1994A&A...282..939H}), 
        especially when images are scaled for optimum display
        contrast.  Type 2 produces dark mottles jutting out from
        network with abrupt endings.  Following the conjecture of De
        Pontieu et al.\ they should show substantial five-minute
        modulation in their Dopplershift and endpoint locations.  The
        bright beginnings of type 2 and 3 mottles that are frequently
        observed near active network and plage are likely due to $S
        \approx \eta B^\star$ Lyman (lines and continuum) irradiation
        by type-1 straws.  Type 1 appears also in on-the-disk \Halpha\
        line center as bright grains and diffuse emission.  In the
        wings only type-2 structures remain visible, always dark
        because the photospheric background $B(T_\rmA)$ is relatively
        bright in \Halpha\
         (\cite{Leenaarts++2006a}). 
        Dopplershifts cause extra type-2 darkening, most frequently in
        the blue wing through blueshift.

  \item \Halpha\ at the limb: types 1 and 2 gain \Halpha\
        conspicuousness in the line wings in which type 3 vanishes.
        Larger line width increases their far-wing visibility on the
        disk but decreases their off-limb contrast as spicules
        (A.G. de Wijn).

  \item Modeling: chromospheric modeling should upgrade to the
        comprehensive VAL/FAL radiation treatment of
        \citet{1981ApJS...45..635V} 
        and
        \citet{1993ApJ...406..319F} 
        which covers the whole gamut from high-density LTE in their
        photospheric bottoms through NLTE and PRD to coronal
        conditions at their $T=10^5$\,K model tops -- but not in 1D
        radial geometry but for chromospheric cylinders and sheets,
        with variation in orientation, density, and temperature,
        embedded in coronal or upper-photosphere surroundings, and
        possessing transition-region sheaths to these and upward
        coronal contact as in types 1 and 2. The next step is to feed
        them mass from below through dynamic pistoning including
        $p$-modes and convective squirting
        (\cite{1955ApJ...121..349B}), 
        and energy from above through conduction.  

  \item Nomenclature: the {\em chromosphere} consists primarily of
        type-3 mottles and fibrils.  They appear as a crowded \Halpha\
        forest at the limb and collectively cause the purple \Halpha\
        + \Hbeta\ flash at the onset of totality for which the
        chromosphere is rightfully named.  Later definitions as
        ``initial outward temperature rise'' or ``layer between the
        temperature minimum and the transition region'' should be
        disavowed.  Regime B, extending over $h \approx 400 -
        1300$~km, is non-purple ``upper photosphere'' or
        ``clapotisphere''
          (\cite{1995heli.conf..151R}). 
        Regime C is coronal.  The {\em transition region} is extremely
        warped, consisting of the regime E parts of type-1 straws plus
        the D-F interface in type 2 mottles plus thin low-emissivity
        D-F sheaths along type 2 and 3 mottles.  As a spherical shell
        it exists only around the fictitious but superbly didactic star
        VALIII
         (\cite{Rutten2003:RTSA}).

  \item Cool-star activity: the superbasal ``\CaII\ emission'' which
        is so useful as stellar activity indicator
          (\cite{1991A&A...252..203R}, 
           \cite{1995A&ARv...6..181S}) 
        comes primarily from type 1 structures, just like \Lyalpha\
        and other UV lines.  That's why \CaII\ \HK\ excess flux
        correlates so well with UV line fluxes.  \HK\ network bright
        points are ``transition region'' diagnostics just like those.

\end{enumerate}

\acknowledgements \small I thank Pit S\"utterlin for the DOT
  observations, Marcel Haas, Jorrit Leenaarts, J\'ulius Koza, John
  Raymond, and Alfred de Wijn for discussions and comments, Han
  Uitenbroek for hospitality, the Leids Kerkhoven-Bosscha Fonds for
  travel support, and NASA's Astrophysics Data System for 
  serving much literature.



\begin{thebibliography}{}

\bibitem[\protect\astroncite{{Babcock} \&
  {Babcock}}{1955}]{1955ApJ...121..349B}
{Babcock} H.~W., {Babcock} H.~D., 1955,
  \apj~  121, 349

\bibitem[\protect\astroncite{{Beckers}}{1968}]{1968SoPh....3..367B}
{Beckers} J.~M., 1968,
  \solphys~  3, 367

\bibitem[\protect\astroncite{{Beckers}}{1972}]{1972ARA&A..10...73B}
{Beckers} J.~M., 1972,
  \araa~  10, 73

\bibitem[\protect\astroncite{{Berger} et~al.}{1999}]{1999SoPh..190..409B}
{Berger} T.~E., {De Pontieu} B., {Fletcher} L., {Schrijver} C.~J., {Tarbell}
  T.~D., {Title} A.~M., 1999,
  \solphys~  190, 409

\bibitem[\protect\astroncite{{Carlsson} \& {Stein}}{1997}]{1997ApJ...481..500C}
{Carlsson} M., {Stein} R.~F., 1997,
  \apj~  481, 500

\bibitem[\protect\astroncite{{Carlsson} et~al.}{2004}]{2004ApJ...610L.137C}
{Carlsson} M., {Stein} R.~F., {Nordlund} {\AA}., {Scharmer} G.~B., 2004,
  \apjl~  610, L137

\bibitem[\protect\astroncite{{De Pontieu} et~al.}{2004}]{2004Natur.430..536D}
{De Pontieu} B., {Erd{\'e}lyi} R., {James} S.~P., 2004,
  \nat~  430, 536

\bibitem[\protect\astroncite{{Fontenla} et~al.}{1993}]{1993ApJ...406..319F}
{Fontenla} J.~M., {Avrett} E.~H., {Loeser} R., 1993,
  \apj~  406, 319

\bibitem[\protect\astroncite{{Giovanelli}}{1967}]{1967AuJPh..20...81G}
{Giovanelli} R.~G., 1967,
  Australian Journal of Physics~  20, 81

\bibitem[\protect\astroncite{{Heinzel} \&
  {Schmieder}}{1994}]{1994A&A...282..939H}
{Heinzel} P., {Schmieder} B., 1994,
  \aap~  282, 939

\bibitem[\protect\astroncite{{Keller} et~al.}{2004}]{2004ApJ...607L..59K}
{Keller} C.~U., {Sch{\"u}ssler} M., {V{\"o}gler} A., {Zakharov} V., 2004,
  \apjl~  607, L59

\bibitem[\protect\astroncite{{Korendyke} et~al.}{2001}]{2001SoPh..200...63K}
{Korendyke} C.~M., {Vourlidas} A., {Cook} J.~W., {Dere} K.~P., {Howard} R.~A.,
  {Morrill} J.~S., {Moses} J.~D., {Moulton} N.~E., {Socker} D.~G., 2001,
  \solphys~  200, 63

\bibitem[\protect\astroncite{Leenaarts et~al.}{2006a}]{Leenaarts++2006a}
Leenaarts J., Rutten R.~J., S{\"u}tterlin P., Carlsson M., Uitenbroek H., 2006,
  ~\aap~ 449, 1209

\bibitem[\protect\astroncite{{Leenaarts} \&
  {Wedemeyer-B{\"o}hm}}{2005}]{2005A&A...431..687L}
{Leenaarts} J., {Wedemeyer-B{\"o}hm} S., 2005,
  \aap~  431, 687

\bibitem[\protect\astroncite{{Mazzotta} et~al.}{1998}]{1998A&AS..133..403M}
{Mazzotta} P., {Mazzitelli} G., {Colafrancesco} S., {Vittorio} N., 1998,
  \aaps~  133, 403

\bibitem[\protect\astroncite{{Rutten} et~al.}{1991}]{1991A&A...252..203R}
{Rutten} R.~G.~M., {Schrijver} C.~J., {Lemmens} A.~F.~P., {Zwaan} C., 1991,
  \aap~  252, 203

\bibitem[\protect\astroncite{{Rutten}}{1995}]{1995heli.conf..151R}
{Rutten} R.~J., 1995,
\newblock in ESA SP-376: Helioseismology,  151

\bibitem[\protect\astroncite{{Rutten}}{1999}]{1999ASPC..184..181R}
{Rutten} R.~J., 1999,
\newblock in ASP Conf. Ser. 184: Third Advances in Solar Physics
  Euroconference: Magnetic Fields and Oscillations, p.~181

\bibitem[\protect\astroncite{Rutten}{2003}]{Rutten2003:RTSA}
Rutten R.~J., 2003,
\newblock Radiative Transfer in Stellar Atmospheres,
\newblock Lecture Notes Utrecht University,
  8th Edition, \url{http://www.astro.uu.nl/~rutten}

\bibitem[\protect\astroncite{{Rutten} et~al.}{2004}]{2004A&A...413.1183R}
{Rutten} R.~J., {Hammerschlag} R.~H., {Bettonvil} F.~C.~M., {S{\"u}tterlin} P.,
  {de Wijn} A.~G., 2004,
  \aap~  413, 1183

\bibitem[\protect\astroncite{{Schrijver}}{1995}]{1995A&ARv...6..181S}
{Schrijver} C., 1995,
  \aapr~  6, 181

\bibitem[\protect\astroncite{{Schrijver} \&
  {Title}}{2003}]{2003ApJ...597L.165S}
{Schrijver} C.~J., {Title} A.~M., 2003,
  \apjl~  597, L165

\bibitem[\protect\astroncite{{Shelyag} et~al.}{2004}]{2004A&A...427..335S}
{Shelyag} S., {Sch{\"u}ssler} M., {Solanki} S.~K., {Berdyugina} S.~V.,
  {V{\"o}gler} A., 2004,
  \aap~  427, 335

\bibitem[\protect\astroncite{{Sheminova} et~al.}{2005}]{2005A&A...437.1069S}
{Sheminova} V.~A., {Rutten} R.~J., {Rouppe van der Voort} L.~H.~M., 2005,
  \aap~  437, 1069

\bibitem[\protect\astroncite{{Vernazza} et~al.}{1981}]{1981ApJS...45..635V}
{Vernazza} J.~E., {Avrett} E.~H., {Loeser} R., 1981,
  \apjs~  45, 635

\end{thebibliography}

\end{document}